\documentclass{aa}  
\usepackage{graphicx}
\usepackage{txfonts}
\begin{document}
   \title{The quiet Sun's magnetic flux estimated from \\ 
Ca\,II\,H bright inter-granular G-band structures} 

   \author{B. Bovelet
%          \inst{1}
          \and
          E. Wiehr
%\inst{2}\fnmsep\thanks{Just to show the usage of the elements 
%in the author field}
          }

   \offprints{E. Wiehr}

   \institute{Institut f\"ur Astrophysik, Universit\"at G\"ottingen,
              Friedrich-Hund-Platz 1, D-37077 G\"ottingen\\
              \email{ewiehr@astro.physik.uni-goettingen.de}
             }

   \date{Received March 4, 2008; accepted June 5, 2008}

% \abstract{}{}{}{}{} 
% 5 {} token are mandatory
 
  \abstract
    {}
   {We determine the number density and area contribution of small-scale 
inter-granular Ca\,{\small II} bright G-band structures in images of the 
quiet Sun as tracers of kilo-Gauss magnetic flux-concentrations.}
   {In a $149\arcsec\,\times\,117\arcsec$ G-band image of the disk center at 
the activity minimum, 7593 small inter-granular structures were segmented with 
the `multiple-level tracking' pattern recognition algorithm. The scatterplot 
of the continuum versus the G-band brightness shows the known magnetic 
and non-magnetic branches. These branches are largely disentangled by 
applying an intrinsic Ca\,{\small II}\,H excess criterion. The thus obtained 
2995 structures contain 1152 G-band bright points (BP) and 1843 G-band 
faint points (FP). They show a tendency toward increasing size
with decreasing G-band excess, as expected from the `hot wall' picture.
Their Ca\,{\small II}\,H and G-band brightness are slightly related, 
resembling the known relation of Ca\,{\small II} and magnetic field 
strength. The magnetic flux density of each individual BP and FP is
estimated from their G-band brightness according to MHD model 
calculations.}
   {The entity of BP and FP covers the total FOV with a number density 
of 0.32\,/\,Mm$^2$ and a total area contribution of 2.0\%. Their 
individual calibrations yield a mean flux density of 20\,Mx/cm$^2$ 
in the entire FOV and 13\,Mx/cm$^2$ for inter-network regions.}
   {}


   \maketitle

%________________________________________________________________

\section{Introduction}
 
The small-scale magnetic flux-concentrations on the non-active solar 
surface are an important ingredient for understanding solar magnetism. 
Active region features, such as sunspots, faculae, prominences, and 
flares, exhibit a strong variation with the solar cycle, commonly 
explained by a global solar dynamo acting in the deep convection 
zone. On the other hand, small-scale magnetic flux-concentrations 
exist that cover the solar surface more homogeneously in time and 
space indicating a less tight dependence on the global dynamo

Unfortunately, it is difficult to measure magnetic flux on those 
scales predicted by MHD modeling (V\"ogler \& Sch\"ussler 2007). 
Measurements of circular polarization due to the Zeeman effect 
can only give the {\it net} flux, since opposite polarities 
cancel each other out within the spatial resolution element 
(Dom\'inguez Cerde\~na et al. 2006). The absolute (unsigned) 
flux, however, has to be deduced from magnetically broadened 
intensity profiles (Stokes-I) that are affected not only by the 
magnetic field but also by the, a priori unknown, intrinsic 
atmosphere of each individual flux concentration. 

The problems for such a conversion have been discussed by, e.g., 
S\'anchez Almeida (2003). Algorithms used for this procedure are 
often called `inversion codes', which might suggest that the observed 
polarization profiles are actually `inverted' into magnetic field 
parameters. Instead, these codes vary both the magnetic field 
structure and the intrinsic atmosphere model until an optimal 
parameter fit between calculated and observed profiles is found. 
The codes also optimize the filling factor of the magnetic signal 
inside the resolution element but do not consider two-dimensional 
radiative effects in very small structures. 

Recent observations 
impressively show that flux-concentrations can be of such a small 
size that even the 0.09\arcsec{} spatial resolution of the 1\,m SST 
at La Palma (Scharmer et al. 2003) does not reach the lower end of 
the size distribution of G-band structures. Stokes images, however, 
cannot achieve such a high spatial resolution, due to additional 
optical elements and longer exposure times. Hence, they even depend 
more on the (doubtful) fill factor than intensity images.

Many of the problems in interpreting polarization measurements 
may be avoided by analyzing intensity images (Sanchez Almeida et 
al. 2004; De Wijn et al. 2005). Small-scale magnetic concentrations 
of kilo-Gauss flux density are known to exhibit excess brightness 
in the CH-band near 430\,nm (G-band) over the neighboring continuum 
(Berger et al. 1995), thus allowing their study in post factum 
reconstructed images of higher spatial and temporal resolution than 
achieved for Stokes measurements. The excess in the G-band brightness 
over the neighboring spectral continuum is a different measure than 
the excess over the mean photospheric G-band level, defining G-band 
bright points (BP; Muller \& Roudier 1984). The latter cover two 
families of structures: (I) inter-granular BP, which are magnetic 
(Berger \& Title 2001) and down-drafting (Langhans et al. 2002), 
and (II) non-magnetic BP, which move upwards with the respective 
granules (same two papers). 

We separate both types of BP using a pattern recognition code, 
(Bovelet \& Wiehr 2007) especially adapted to detect the 
inter-granular BP. We then select magnetic inter-granular structures 
using their G-band-to-continuum ratio and additionally their 
Ca\,{\small II}\,H excess brightness. There is some controversy about 
these criteria: Rezaei et al. (2007) find that the Ca\,{\small II} 
excess cannot be considered a general indicator for magnetic flux 
concentrations. Berger et al. (1998) argue that the G-band brightness 
is not a sufficient criterion for the magnetic nature of a BP. De\,Wijn 
et al. (2005) suggest the Ca\,{\small II} brightness to be a more 
realistic criterion. Their concern about the G-band criterion refers 
to Berger \& Title (2001), who, however, only exclude that criterion for 
those BP that appear embedded in granules, i.e., the above family\,II. 

We avoid these, restricting our selection to {\it small} structures, 
thus excluding extended granular features and, hence, non-magnetic 
local G-band enhancements embedded. Among the small-scale inter-granular 
structures (IgS), we remove those that populate the non magnetic branch 
in the G-band-to-continuum scatterplot. The remaining IgS are than 
sampled by a significant Ca\,{\small II} excess. 

The powerful pattern recognition algorithm, followed by a selective 
sampling of small structures, elimination of non-magnetic ones by 
their G-band-to-continuum ratio, and classification of magnetic ones
by the Ca\,{\small II} excess, qualifies our study as an {\it alternate 
way} to investigate small-scale solar kilo-Gauss magnetic flux-concentrations 
at the highest possible spatial resolution and independent of interpreting 
polarimetric measurements.

%___________________________FIGURE-1___________________ 
   \begin{figure}
   \centering
   \includegraphics[width=0.95\linewidth,angle=0]{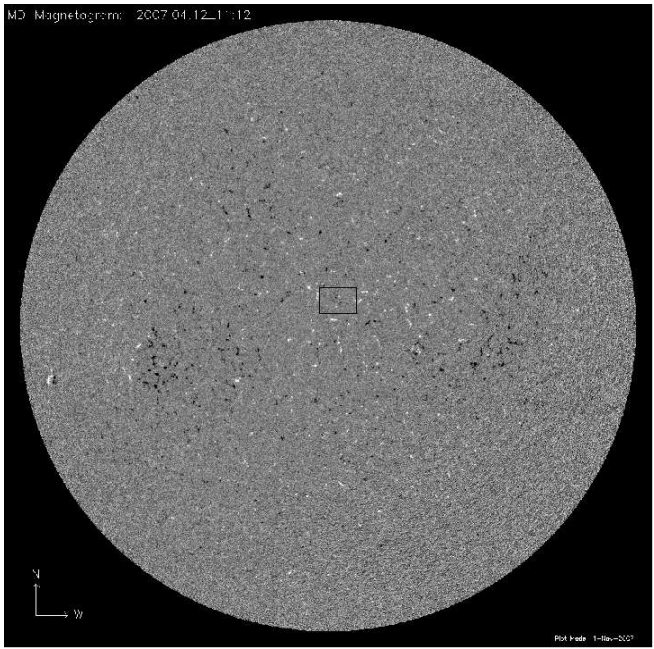}
   \caption{Magnetic map of the solar disk on April 12, 2007 from 
    MDI showing an extremely low level of solar activity. The 
    $149\arcsec\,\times\,117\arcsec$ FOV investigated in this paper is marked.}
              \label{FigMagnetSun}%
    \end{figure}
%_______________________________________________________________________

\section{Segmentation of G-band features}
 
\begin{figure}
   \centering
   \includegraphics[width=0.95\linewidth,angle=0]{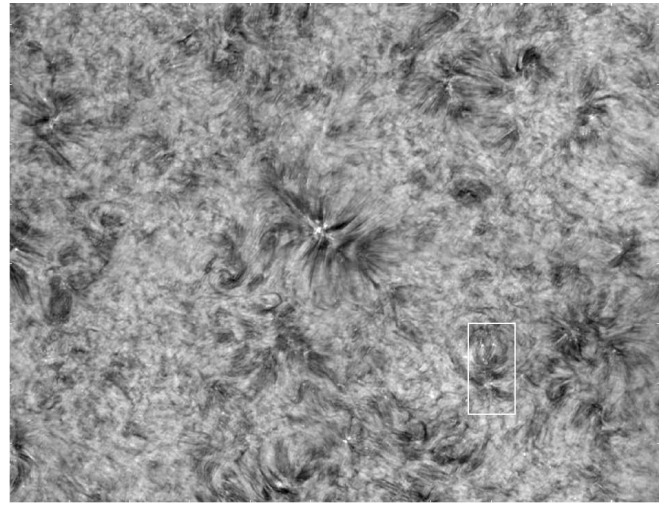}
   \caption{Reconstructed H$\alpha$ image of a 
$149\arcsec\,\times\,117\arcsec$ region centered on the central 
meridian and the equator ($6^o$N) on April 12, 2007, (blue and 
red wing images added; courtesy P. S\"utterlin, Utrecht). 
The square marks a subregion used for a visualization of the 
data processing in Fig.\,\ref{FigSubregion}.}
              \label{FigHalpha}
    \end{figure}

We analyzed a set of high-resolution images, observed at an 
exceptionally deep solar activity minimum on April 12, 2007  
(cf., MDI-magnetogram Fig.\,\ref{FigMagnetSun}). The Dutch Open 
Solar Telescope (DOT) was pointed at the intersection of the 
central meridian with the solar equator (0$^o$\,E/W; 6$^o$\,N) 
by P. S\"utterlin, who took image bursts simultaneously in the 
G-band, Ca\,{\small II}\,H, H$\alpha$, blue and red continuum, 
which he subjected to image reconstruction following de\,Boer 
(1996) and finally composed to a 4-image mosaic of 
$149\arcsec\,\times\,117\arcsec$ total field-of-view. 
That large FOV, marked in Figure\,\ref{FigMagnetSun} and shown in 
Figure\,\ref{FigHalpha} as an H$\alpha$ image, covers 1/221 of 
the total solar surface, allowing a statistically significant 
investigation of small-scale magnetic flux-concentrations. 
 
\begin{figure}[h]
   \centering
   \includegraphics[width=18.0cm,angle=270]{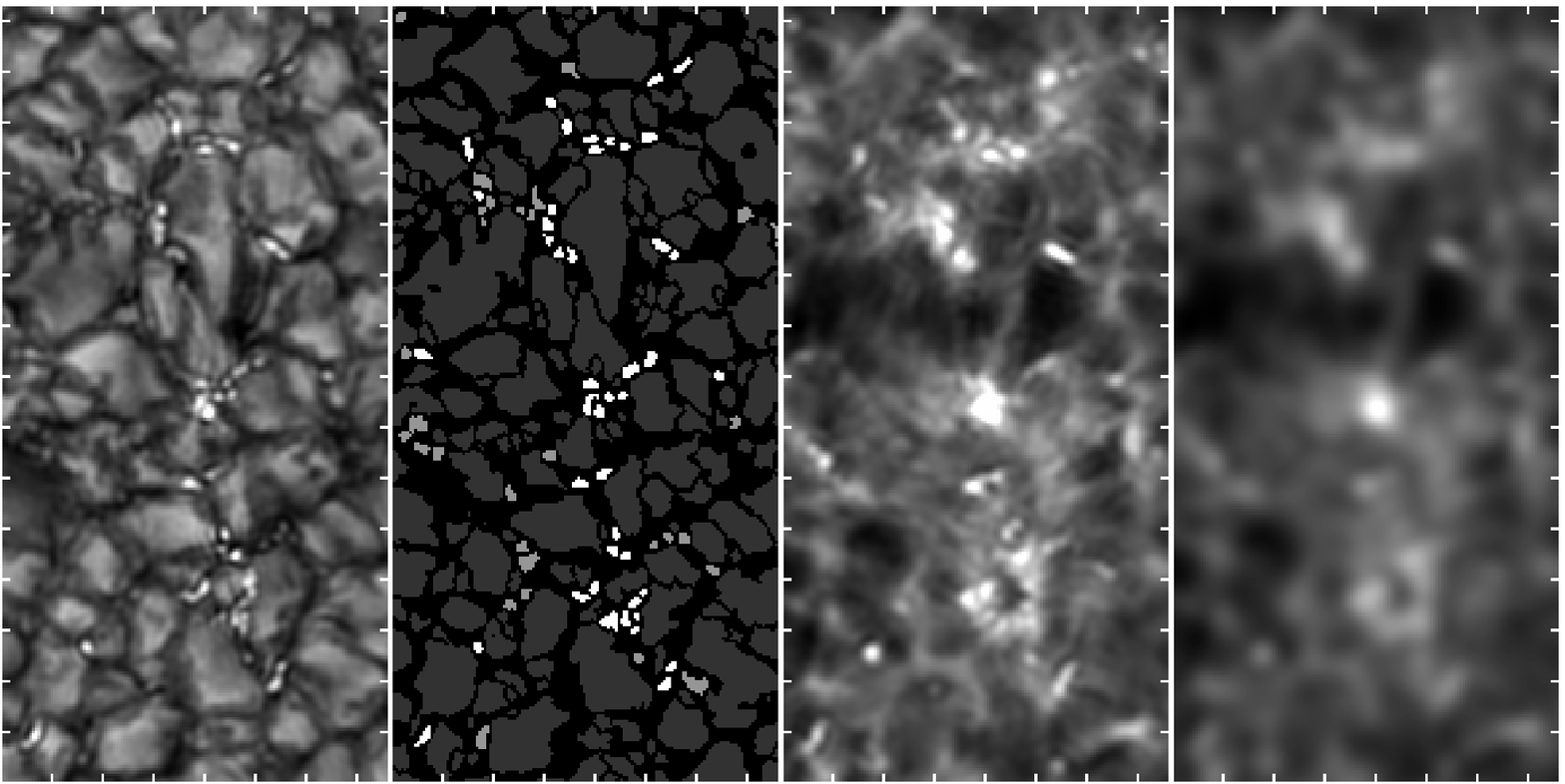}
\caption{Sub-region of $10.4\arcsec\,\times\,20.7\arcsec$ marked in 
Fig.\,\ref{FigHalpha} and rotated by $-90^o$ in the light of the 
G-band, together with the corresponding patterns of bright points 
(BP; white), faint points (FP; light gray), and granules (dark gray); 
the Ca\,{\small II}\,H image without and with artificial degradation 
to 0.75\arcsec{} spatial resolution; tickmarks: 1\,Mm.}
         \label{FigSubregion}
   \end{figure}

As a {\it first step}, the G-band image was analyzed using the MLT\_\,4 
recognition code\footnote{http://www.gwdg.de/$\sim$bbovele ; login 
as `user', password `mlt4'} (Bovelet \& Wiehr 2007) to obtain an 
appropriate pattern of solar structures and their individual sizes 
in pixel counts. The latter depend to some extent on the `unitary cut 
level' equally applied to each of the segmented cells: the higher the 
cut level, the smaller the size. Our final choice of a unitary cut 
level of 0.5 was adapted to a realistic representation of IgS sizes, 
accepting slightly broadened inter-granular lanes. As a {\it second step}, 
we selected inter-granular structures ('IgS') with an area $\le75$ 
pixels ($2\,\times\,10^4$cm$^2$). This procedure excludes larger features 
with local G-band enhancements occasionally embedded that are known 
to be non magnetic (Berger \& Title 2001; Langhans et al. 2002). 
The suitability of the achieved representation is demonstrated in 
Figure\,\ref{FigSubregion} for a sub-field indicated in 
Figure\,\ref{FigHalpha}.

%___________________________FIGURE-2___________________ 
%_______________________________________________________________________

\section{Selection of IgS with Ca\,{\small II}\,H excess}

\begin{figure}[h]
   \centering
   \includegraphics[width=0.95\linewidth,angle=0]{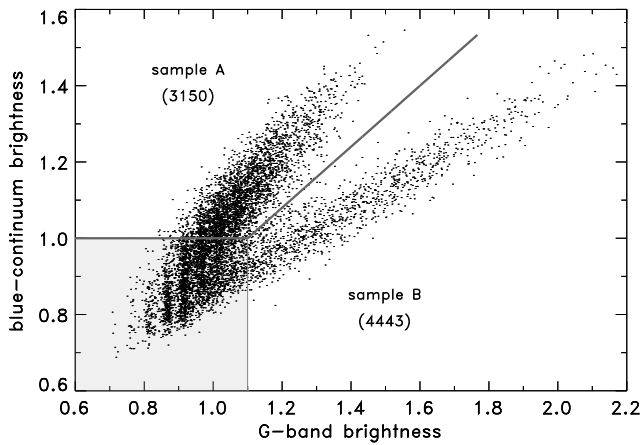}
      \caption{Scatterplot of blue-continuum and G-band brightness 
for all 7593 IgS selected, grouping in a magnetic (lower) and a 
non-magnetic branch that are separated by the full line; overlap 
regime shaded.}
         \label{FigScatterplotBcGb}
   \end{figure}

For the thus selected total of 7593 IgS, we plot the G-band 
excess over the intensity in the neighboring spectral continuum 
following Berger et al. (1998). The spatial correlation between the 
two simultaneously observed G-band and (blue) continuum images is 
optimized by a cross-correlation to one pixel (0.071\arcsec{}). 
Figure\,\ref{FigScatterplotBcGb} shows a scatterplot of the 33\% highest 
intensity pixels of each IgS in the continuum and in the G-band. 
The 33\% restriction (rather than its mean value) ensures that the 
deduced IgS brightness is largely independent of individual shapes. 
Figure\,\ref{FigScatterplotBcGb} gives both brightness values in units 
of the mean photosphere at FOV locations free of local enhancements 
(e.g., the lower-left region in Fig.\,\ref{FigHalpha}). 

The scatterplot in Figure\,\ref{FigScatterplotBcGb} shows two 
branches that, according to measurements by Berger et al. 
(1998) and to calculations by Shelyag et al. (2004), represent 
magnetic and non-magnetic IgS and overlap for G-band brightness 
$I^{Gb}\!<1.1\,\times\,I^{Gb}_{mean}$ (shaded in 
Fig.\,\ref{FigScatterplotBcGb}). Since we do not, a priori, 
know which of the segmented features are magnetic, we remove, 
as a {\it third step}, 3150 non-magnetic IgS in the upper part 
of the non magnetic branch (sample\,A), located above the kinked 
line line inserted in Figure\,\ref{FigScatterplotBcGb}.

%---------------------------FIGURE-3---------------------------
%_____________________________________________________________
 
Among the remaining 4443 IgS (sample\,B), we disentangle 
non-magnetic and magnetic ones by their Ca\,{\small II}\,H 
excess brightness, assuming magnetic IgS to exhibit a substantial 
Ca\,{\small II}\,H excess, whereas non-magnetic IgS are considered 
to have no significant Ca\,{\small II}\,H brightening (Lites et al. 
1999; Rezaei et al. 2007). We correspondingly remove, as a {\it 
fourth step}, the non magnetic IgS from sample\,B applying a 
threshold of the intrinsic Ca\,{\small II}\,H excess defined as 
$$ExCa\,:=\,I_{up}^{\quad{}(I_{up}/I_{low})},$$ 
 
\noindent
with $I_{up}$ denoting the Ca\,{\small II}\,H brightness contrast 
over the mean for 33\% of the brightest and $I_{low}$ for 33\% 
of the darkest pixels of each IgS. Application of the ratio 
of both contrasts as the exponent to $I_{up}$ avoids excluding 
features with a high but flat intrinsic Ca\,{\small II}\,H brightness 
distribution, as G-band bright points. Such a criterion mainly 
selects IgS with a {\it spatially isolated Ca\,{\small II}\,H excess 
inside their corresponding G-band pattern} and excludes faint and 
`fuzzy' Ca\,{\small II}\,H structures (widely considered to have an 
acoustic nature).  

%-------------------------FIGURE-4----------------------------
   \begin{figure}[h]
   \centering
   \includegraphics[width=6.7cm]{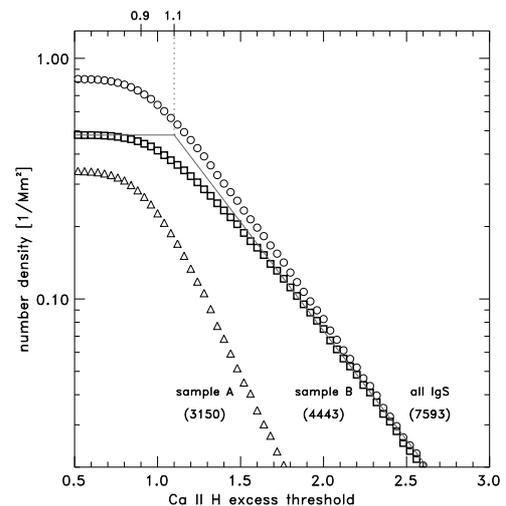}
      \caption{Number density (logarithmic) as a function of 
the intrinsic Ca\,{\small II}\,H excess for all 7593 inter-granular 
structures, the 3150 above and the 4443 below the division line 
in Fig.\ref{FigScatterplotBcGb}; flat top and steep decrease 
approximated by lines intersecting at ExCa=1.1.}
         \label{FigExCaThreshold}
   \end{figure}
%_____________________________________________________________

The influence of this intrinsic Ca\,{\small II}\,H excess criterion on 
the number density of IgS is demonstrated in Figure\,\ref{FigExCaThreshold}. 
For sample\,A, the logarithmic number density of IgS already drops at 
very low ExCa values, since this group is mainly populated with IgS from 
the non magnetic branch without intrinsic Ca\,{\small II}\,H excess. They 
mark granular fragments or faint and `fuzzy' Ca\,{\small II}\,H brightenings 
that are already removed by a low $ExCa$ value, since both basis and exponent 
of $ExCa$ are small. 

%------------------------FIGURE-5----------------------------
%_____________________________________________________________

%---------------------------FIGURE-6-------------------------------
   \begin{figure*}
   \centering
   \includegraphics[width=\linewidth]{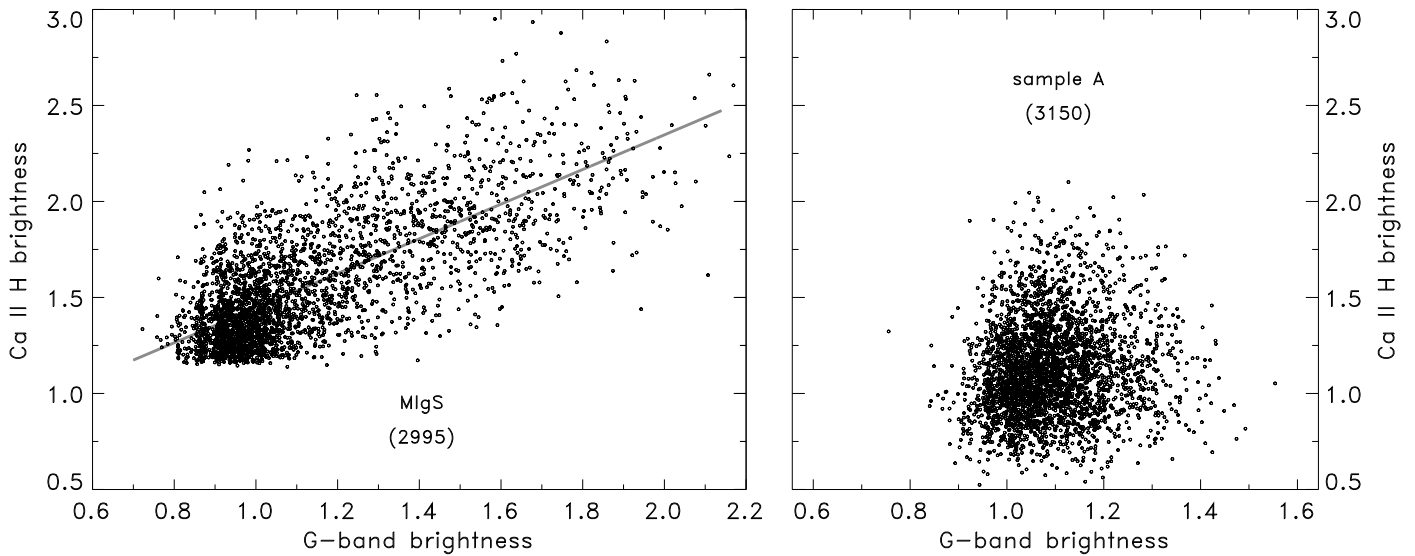} 
      \caption{Scatterplot of the brightness in Ca\,{\small II}\,H and 
in the G-band for the 2995 Ca\,{\small II}\,H bright IgS from sample\,B 
(MIgS) and for the 3150 IgS from sample\,A (right panel), showing a 
tight relation only for the MIgS (here linearly fitted); the lower 
limit of the ordinate values marks the influence of the Ca\,{\small II}\,H
threshold applied to sample\,B .} 
         \label{FigScatterplotCaGb}
   \end{figure*}
%_____________________________________________________________

For sample\,B, the logarithmic number density of IgS remains constant 
up to $ExCa\!\approx\!0.9$, then drops, and finally linearly decreases 
for $ExCa\!>\!1.4$. The high Ca\,{\small II}\,H excess of IgS in sample\,B 
causes already low $ExCa$ thresholds to exclude most of the non-magnetic 
IgS, then leaving the number density constant. When further increasing 
the threshold to $ExCa\!>\!0.9$, magnetic IgS with Ca\,{\small II}\,H 
excess are also excluded progressively. 

We eventually eliminate from sample\,B those IgS that do not reach 
$ExCa\!=\!1.2$, and obtain 2995 remaining IgS, which are 
Ca\,{\small II}\,H bright and thus reasonably assumed to be magnetic, 
hence, denoted as `MIgS'. Among these, G-band bright MIgS with 
$I^{Gb}\!\ge\!1.1\,\times\,I^{Gb}_{mean}$ (i.e. beyond the overlap 
regime) are the known BP. Accordingly, we denote the remaining MIgS 
with $I^{Gb}\!<1.1\,\times\,I^{Gb}_{mean}$ as `G-band faint points' 
(FP).

\section{Results}

\subsection{Ca\,{\small II} brightening}
 
Figure\,\ref{FigSubregion} gives the visual impression that the 
Ca\,{\small II}\,H brightness is not tightly related to the G-band 
brightness. This is quantitatively seen in the scatterplots in
Figure\,\ref{FigScatterplotCaGb}. Our 2995 MIgS show a fairly good 
relation that resembles the one found by Rezaei et al. (2007) 
between the Ca\,{\small II}\,K index and the magnetic flux density.
The smooth transition between BP and FP (at 
$I^{Gb}\!=\!1.1\,\times\,I^{Gb}_{mean}$) indicates their similar
(magnetic) nature, as found by Bovelet \& Wiehr (2007).
The absence of such a relation for the 3150 IgS from sample\,A
(right panel) indicates that these are mainly not (or weak) 
magnetic. 

\begin{figure}[h]
   \centering
   \includegraphics[width=0.9\linewidth,angle=0]{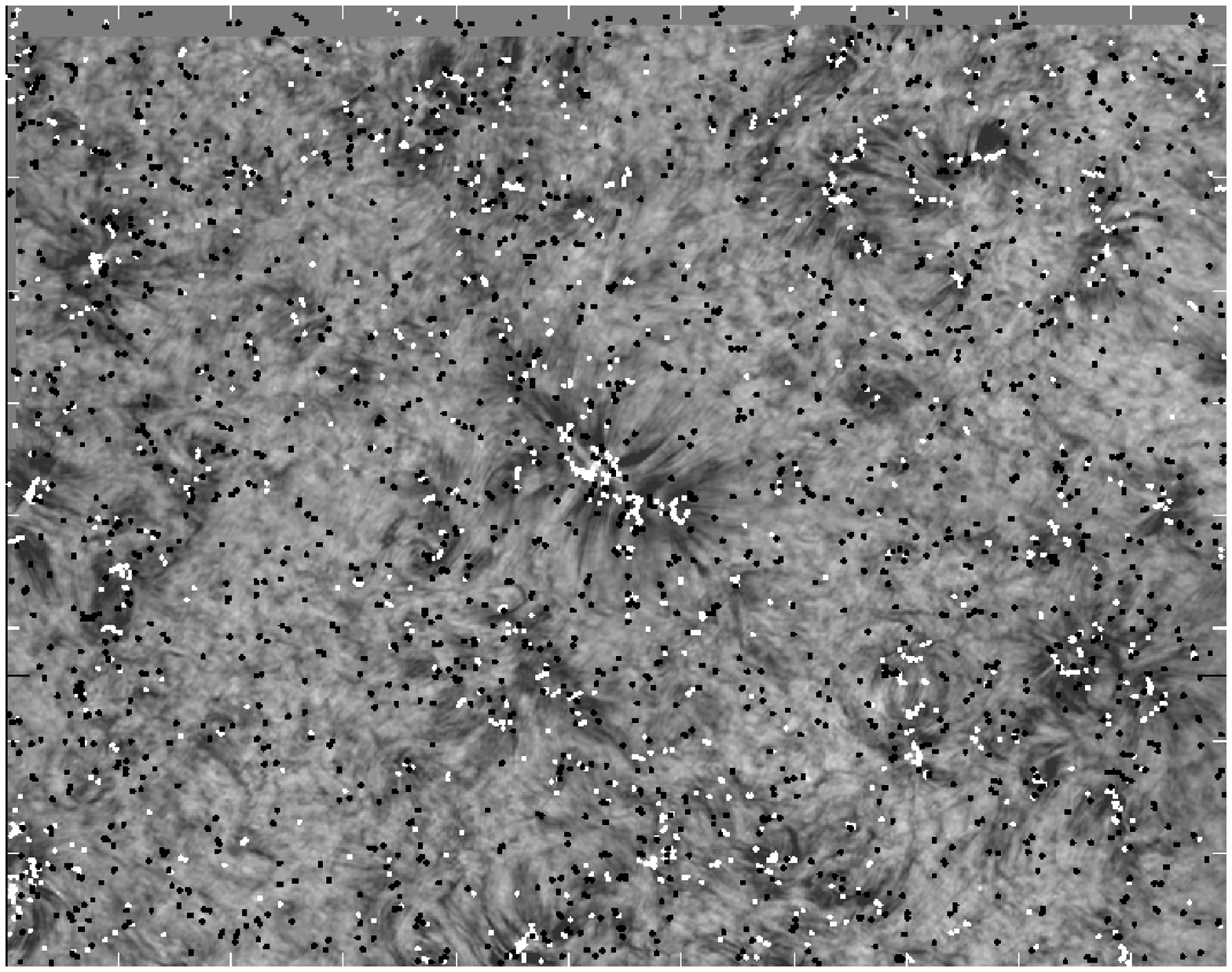}
   \caption{Location of magnetic inter-granular structures MIgS
within the H$\alpha$ structures of the $149\arcsec\!\times\!117\arcsec$ FOV;
white dots mark G-band bright MIgS (BP), black dots mark faint ones 
(FP); the dots represent locations rather than feature sizes; gray 
peripheral zones not covered by the H$\alpha$ mosaic; tickmarks: 10\,Mm.}
         \label{FigSGpattern}
   \end{figure}

The findings by Lites et al. (1999) and by Rezaei et al. (2007) that 
magnetic flux concentrations in the network cell interior do not 
exhibit a Ca\,{\small II} excess might suggest that most of our MIgS 
are located at network boundaries. To check this idea, we superpose 
the H$\alpha$ image from Figure\,\ref{FigHalpha} with the location 
of our finally selected MIgS (Fig.\,\ref{FigSGpattern}). This 
superposition impressively shows that the BP are preferably located 
near the dark H$\alpha$ fibrils, which are widely assumed to have 
their roots in small-scale kilo-Gauss magnetic flux-concentrations. 

%-------------------------FIGURE-7-----------------------------
%
%_____________________________________________________________

In contrast, FP appear also outside the H$\alpha$ fibrils, i.e., in 
the inter-network (IN). This, however, means that {\it at least part 
of the IN flux-concentrations do exhibit marked Ca\,{\small II}\,H 
excess}, since the FP have been equally selected by the $ExCa$ threshold. 
The discrepancy of this finding to other observations might be due to 
the higher spatial resolution achieved in our data. This is readily 
seen in the lowest panel of Figure\,\ref{FigSubregion}, which displays 
the same region as the upper panels but after artificial degradation 
of the Ca\,{\small II} image simulating a spatial resolution of 
0.75\arcsec{} (15\,cm telescope aperture). It impressively shows, 
that even at such moderate degradation, the isolated tiny 
Ca\,{\small II}\,H bright features are no longer detectable. Our 
intrinsic Ca\,{\small II} excess criterion even fails to select 
IgS for a faint degradation to a 0.5\arcsec{} spatial resolution. 
Hence, {\it very high spatial resolution is required to check 
whether small-scale magnetic flux-concentrations actually exhibit 
a Ca\,{\small II} excess}.

\subsection{Size}
 
\begin{figure}[h]
   \centering
   \includegraphics[width=\linewidth,angle=0]{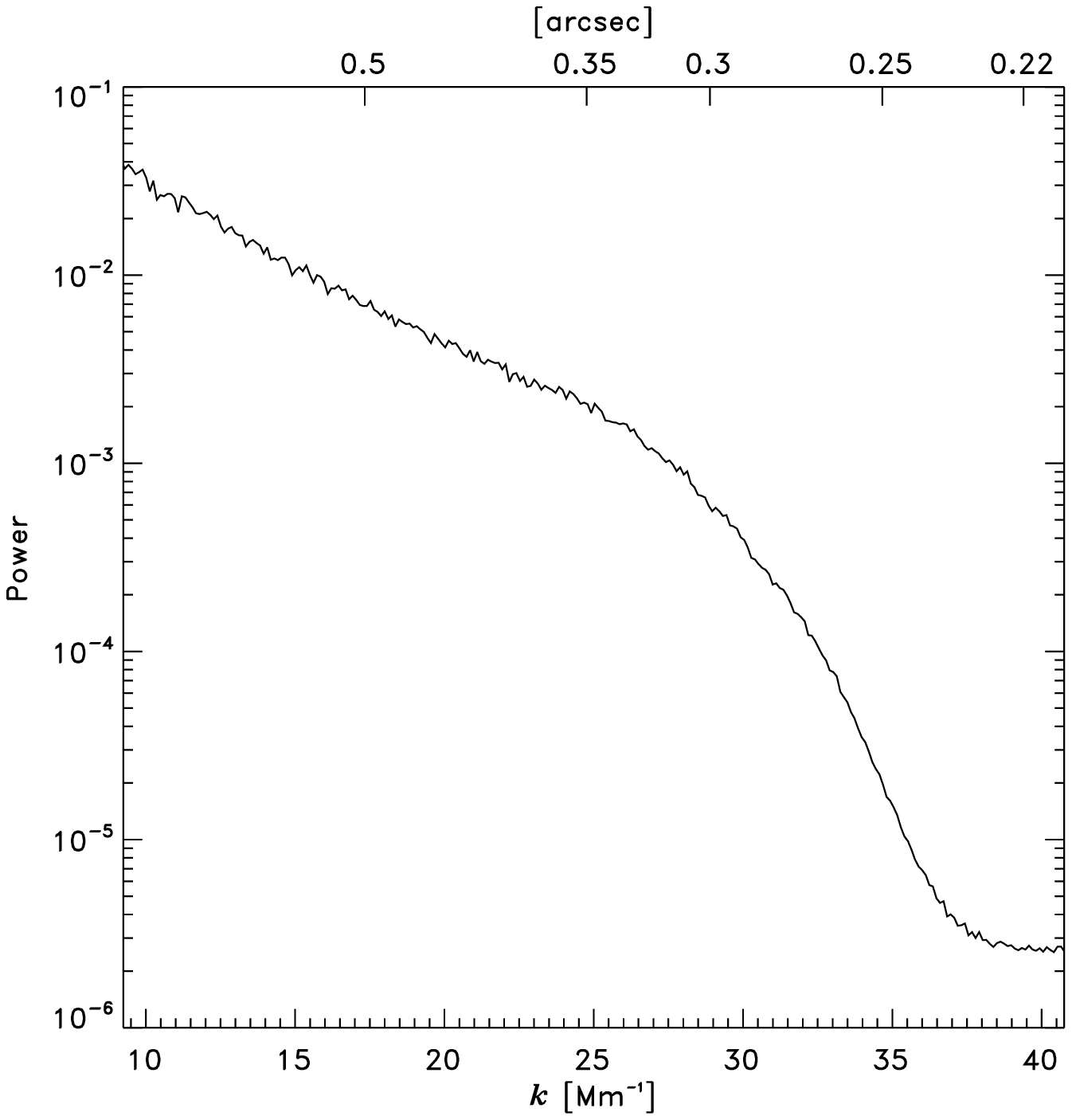}
      \caption{Power spectrum of the G-band image showing a 
limiting resolution of $2\pi\,\times\,k =37$/Mm of the DOT mosaic 
image observed and reconstructed by P. S\"utterlin.}
         \label{FigPowerSpectrum}
   \end{figure}

\begin{figure}[h]
   \centering
   \includegraphics[width=\linewidth,angle=0]{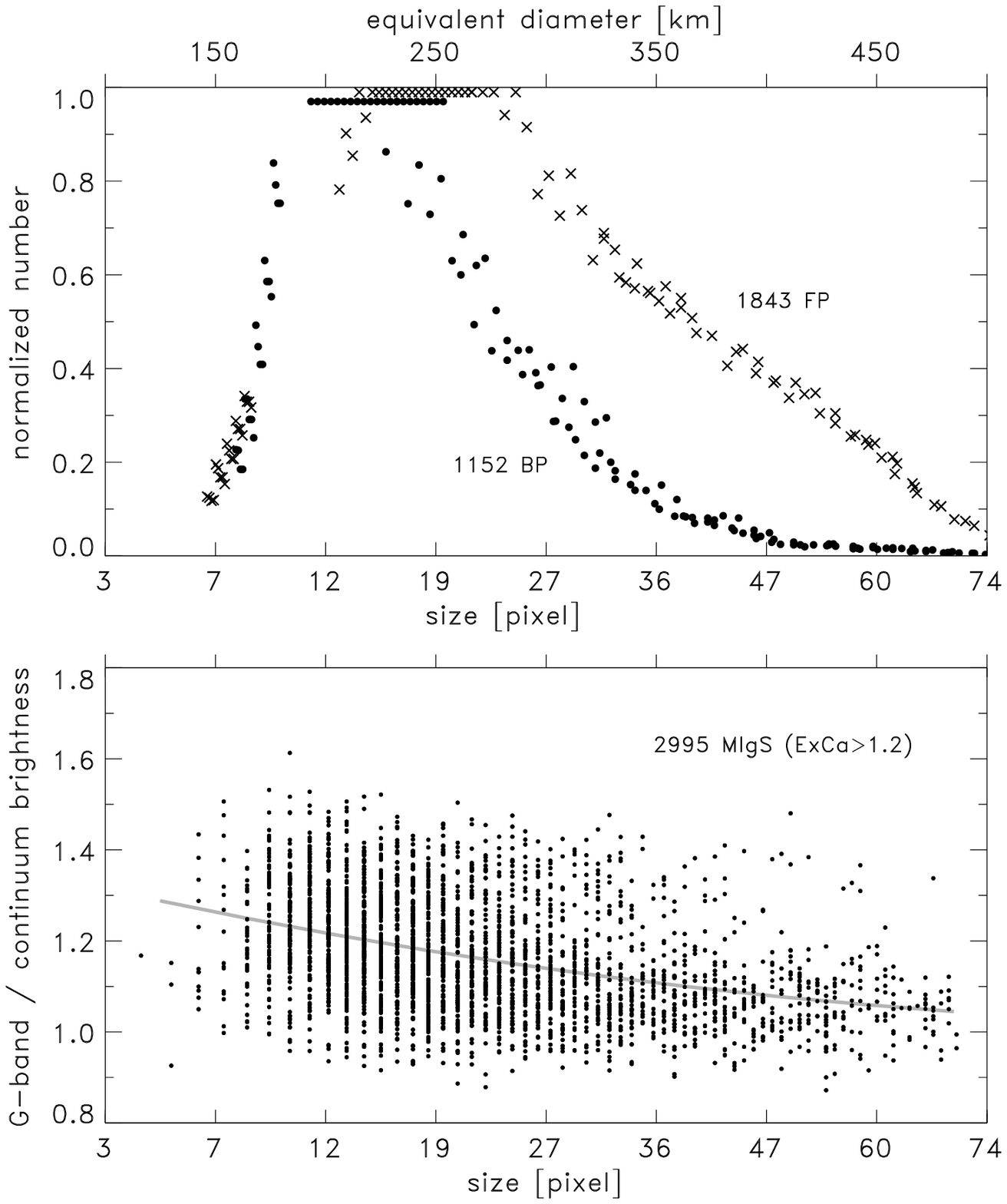}
      \caption{Size histograms of 1152 BP and 1843 FP from sample\,B. 
The lower abscissa gives realistic sizes in pixel counts; the upper 
abscissa gives equivalent diameters of corresponding circular 
shapes. The data points represent counts per bin-size of 21 
maximum-scaled histograms with different bin-sizes that are 
overplotted to provide enough invariance from subjective sampling. 
The lower panel shows the G-band brightness excess (with respect 
to the neighboring spectral continuum) as a function of the size 
for all 2995 MIgS (BP plus FP).}
         \label{FigSizeHisto}
   \end{figure}

The spatial resolution achieved in our data is proven by the power 
spectrum (Fig.\,\ref{FigPowerSpectrum}), which drops to twice the noise 
level at $2\pi\!\cdot\!k\approx\!37$\,Mm$^{-1}$, i.e., 170\,km on the Sun. 
Accordingly, the histograms of BP and FP in Figure\,\ref{FigSizeHisto}a 
drop towards small sizes, reaching 170\,km (0.23\arcsec{}) at 5\% of their 
maxima. The most frequent sizes of the 1152 BP range between 12 and 
20 pixels, those of the 1843 FP between 14 and 23. Since our analysis 
does not need any assumption about feature shapes, we only give diameters 
for equivalent circular shapes additionally at the upper abscissa 
of Figure\,\ref{FigSizeHisto}a and obtain most frequent circular 
diameters of $(220\pm30)$\,km for BP and $(250\pm30)$\,km for FP.

%------------------------FIGURE-8------------------------------
%_____________________________________________________________

Towards increasing sizes, the histogram of BP establishes the 
existence of an upper size limit for the BP (cf., Wiehr, Bovelet \&
Hirzberger 2004; Puschman \& Wiehr 2006; Beck et al. 2007). The 
corresponding dia\-meter of an equivalent circular shape amounts to 
400\,km, in agreement with Muller (1983) and with Bovelet \& Wiehr 
(2003). The FP seem to be systematically larger. We verify this by 
plotting the G-band brightness excess of the MIgS over the neighboring 
spectral continuum as a function of the size and find that {\it 
larger MIgS are primarily faint} (Fig.\,\ref{FigSizeHisto}b). 

This is a hint that for larger magnetic flux-concentrations the `hot 
wall' is wider, thus reducing the brightness excess. In turn, small FP 
do exist that are of similar size as the BP (lower left part of the 
data cloud in Figure\,\ref{FigSizeHisto}b); their `closer hot walls' 
must then be fainter, probably due to a lower magnetic flux density 
(cf., 4.4). The few data points in the upper right of 
Figure\,\ref{FigSizeHisto}b represent 14 elongated MIgS (among 2995)
that have not been segregated enough by MLT\_\,4; this proves the 
power of our algorithm.

\eject

%----------------------- FIGURE-9-----------------------------
%_____________________________________________________________

%----------------------- FIGURE-10-----------------------------
   \begin{figure*}
   \centering
   \includegraphics[width=0.9\linewidth,angle=0]{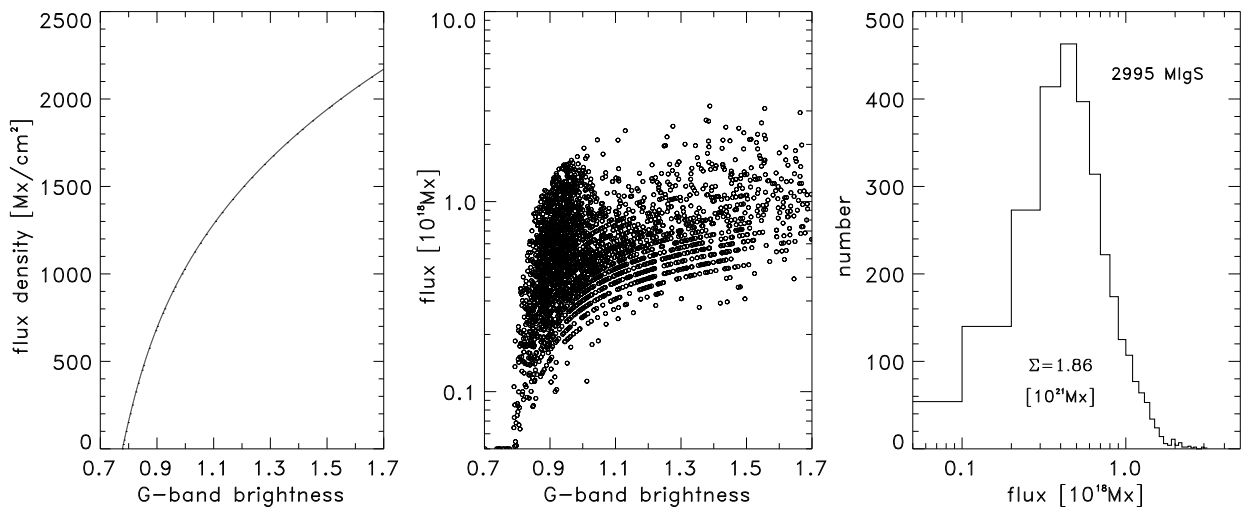}
       \caption{Calibration curve for MIgS (left) averaged 
from Fig.\,4 in Shelyag et al. (2004); scatterplot of 
G-band brightness and magnetic flux (middle) and flux 
histogram (right panel) of 2995 MIgS.}
         \label{FigFLUX}
   \end{figure*}
%_____________________________________________________________

\subsection{Area contribution}

We reasonably assume that the G-band appearance of each MIgS is a 
realistic measure of the corresponding kilo-Gauss magnetic feature, 
following Shelyag et al. (2004), who state that the ``spatial G-band 
pattern matches the spatial structure of the magnetic feature''. Hence, the 
sum of all 2995 MIgS sizes may be considered as an actual area contribution 
in the $9.2\,\times\,10^{19}\!$cm$^2$ FOV ($3.48\,\times\,10^6$\,px). The 
2995 MIgS cover $1.85\,\times\,10^{18}\,$cm$^2$ (69719 px) corresponding 
to 2.0\% of the FOV and a number density of 0.32\,Mm$^{-2}$ (Table\,1), 
i.e., the same as found by S\'anchez Almeida et al. (2004).  

The location of BP and FP in the H$\alpha$ image (Fig.\,\ref{FigSGpattern})
allows to distinguish their affiliation to fibril and non-fibril (IN) regions. 
The `voids of fibrils' appear to be homogeneously populated by FP, whereas 
the BP primarily occur in centers of fibrils regions, as is expected from 
their nature as magnetic footpoints. For a quantitative analysis, we 
approximated those complementary regions by a binary mask, obtained from
sufficient smoothing of the H$\alpha$ image. We find a number density of 
0.22\,Mm$^{-2}$ in the fibril voids, i.e., about 2/3 of the 0.32\,Mm$^{-2}$ 
found for the entire FOV (Table\,1). We verified that this value does not 
significantly depend on the smoothing for a binary mask.

\subsection{Estimate of magnetic parameters}
 
To obtain the mean magnetic flux density of the MIgS in the FOV, we 
might multiply their area contribution of 2\% by a `typical' value 
for an intrinsic flux density of 1\,kG and obtain a mean (unsigned) 
flux density of 20\,G. A possible way estimating the intrinsic flux 
of each MIgS may be obtained from the relation between the G-band 
brightness and the magnetic flux density, as modeled by Shelyag et al. 
(2004). There, in Fig.\,4, most of the scatter of the G-band brightness 
arises from the spatially varying continuum level in their calculation 
area (also containing granules). Our sample of MIgS, however, is 
restricted to small inter-granular features with a certain G-band 
excess and will be less influenced by continuum variations. We may
thus consider the mean MHD relation as calibration.
 
The modeled relation refers to each pixel of a magnetic feature. 
We averaged the G-band brightness of the pixels of each MIgS and 
used that mean for calibration (left panel of Fig.\,\ref{FigFLUX}), 
thus obtaining the intrinsic magnetic flux density. Multiplying 
this value with the corresponding area (returned by MLT\_\,4) 
yields a flux value for each MIgS. The middle panel of 
Figure\,\ref{FigFLUX} shows these flux values as a function 
of the mean G-band brightness for each of our 2995 MIgS. 

The flux values depend to some extent on the unitary cut level
(0.5; cf., Section\,2) applied to each segmented cell after 
normalization to its brightness maximum in phase\,4 of MLT\_\,4 
(cf., Bovelet and Wiehr 2007): the higher the cut level, the 
smaller the size. Choosing a lower cut threshold would expand 
MIgS to their peripheral pixels that are fainter and thus would 
lower the mean G-band brightness of the features. The calibration 
curve (left panel of Fig.\,\ref{FigFLUX}) will then assign lower 
flux densities, and the flux of this expanded MIgS will increase 
by only a few percent. The influence of our unitary cut level on
the deduced MIgS flux is thus negligible.

\subsection{Mean flux in the FOV}

The histogram of the flux of the MIgS (right panel of 
Fig.\,\ref{FigFLUX}) shows a narrow distribution with most 
frequent values of $0.45\,\times\,10^{18}$\,Mx and a total 
flux of $1.86\,\times\,10^{21}$\,Mx for all 2995 MIgS. If we 
divide the latter value by the total area of the 2995 MIgS 
($1.85\,\times\,10^{18}$cm$^2$), we obtain a {\it most 
frequent MIgS flux density} of 1005\,Mx/cm$^2$, showing 
that the kilo-Gauss assumption is a realistic mean over 
the entire FOV. But the most frequent flux density largely 
differs between BP and FP, amounting to 1597\,Mx/cm$^2$ for 
the BP and to 775\,Mx/cm$^2$ for the FP.

The difference between BP and FP is not that great, however,
for the {\it mean flux density in the entire FOV}. Dividing the 
total flux of the 1152 BP and that of the 1842 FP separately 
by the entire FOV area, we obtain 9.6\,Mx/cm$^2$ for the BP and 
10.6\,Mx/cm$^2$ for the FP (which are more frequent but less 
magnetic, see above). The sum of both values gives a mean flux 
density over the FOV of 20.3\,Mx/cm$^2$ for all 2995 MIgS, 
a number that ranges between that by Daras-Papamargaritis \& 
Koutchmy (1983; $10^{21}$\,Mx/$10^{20}$cm$^2$) and the one by
S\'anchez Almeida (2003; 30\,Mx/cm$^2$ for his full FOV).

% ----------------------One column table--------------------------
   \begin{table}[h]
\caption[]{Characteristics of IgS with $ExCa\,>\,1.2$, distinguished 
for BP ($I^{Gb}\!\ge 1.1\,\times\,I^{Gb}_{mean}$), FP  
($I^{Gb}\!<1.1\,\times\,I^{Gb}_{mean}$), and MIgS (BP plus FP) 
in the entire FOV; MIgS also in regions without pronounced 
H$\alpha$ fibrils.}

\label{table:1}      % is used to refer this table in the text
\centering                          % used for centering table
\renewcommand{\arraystretch}{1.15}  % increases line feed
\begin{tabular}{c c c c c}        % centered columns (5 columns)
sampled & total  & no.~dens.   & area~cover. & mean~flux~dens.\\
entity  & no. & [Mm$^{-2}$] & [$\%$] & [Mx/cm$^2$]  \\[0.2ex]
\hline                               
BP        &  1152 & 0.12 & 0.60 & \,\,9.6\\[0.2ex]
FP        &  1843 & 0.20 & 1.40 & 10.7\\[0.2ex]
\hline             % inserts horizontal line
MIgS in FOV  &  2995 & 0.32 & 2.00 & 20.3\\[0.2ex]
MIgS in voids &   888 & 0.22 & 1.52 & 12.8\\[0.2ex]
\end{tabular}
\end{table}
\renewcommand{\arraystretch}{1.0}   % resets line feed

%_____________________________________________________________

In the `fibril voids' seen in Figure\,\ref{FigHalpha}, we find 
0.22\,MIgS/Mm$^2$ covering 1.52\% of the corresponding void area 
(Table\,1). Assuming kilo-Gauss field strengths for each MIgS, we 
obtain a mean flux density in the fibril voids of 15.2\,Mx/cm$^2$, 
in perfect agreement with the finding by S\'anchez Almeida 
(2003; for the network interior). However, attributing to
each MIgS its individual magnetic flux via the calibration 
curve (left panel of Fig.\,\ref{FigFLUX}) yields a mean flux 
density of only 12.8\,Mx/cm$^2$ for the void areas. This value 
is lower than the above one for kilo-Gauss fields, since the 
fibril voids are primarily populated by FP, which exhibit 
lower most frequent flux densities. We verified that the 
reduction of the mean flux density in fibril voids as compared 
to that in the entire FOV does not significantly depend on the 
binary mask used to exclude fibril regions (cf., Table\,1).

\section{Conclusions}
 
The segmentation and selection of small-scale inter-granular structures 
with a significant Ca\,{\small II}\,H excess allows lower limits to be 
determined for the number density and fractional area coverage of 
inter-granular kilo-Gauss magnetic flux-concentrations. A magnetic 
calibration of each feature yields a mean (unsigned) magnetic flux 
density in a quiet solar region close to existing results. Our method 
offers an alternate approach to polarimetry, although it inevitably misses 
those magnetic structures that do not exhibit a Ca\,{\small II}\,H excess 
and `elusive magnetic structures' from fluctuations in thermodynamic 
properties (which S\'anchez Almeida, 2000, estimate to carry at least 
half of the solar magnetic flux). 

In case the total flux in our disk center FOV were representative 
of the entire solar surface ($221\,\times\,$FOV), the Sun would be
covered by $>\!0.6\!\times\!10^6$ MIgS with a total flux of
$>\!4.1\,\times\,10^{23}$\,Mx. Considering also magnetic structures
which are not covered by our selection, the Sun at activity 
minimum may well be covered by a few millions of MIgS with a 
total flux of $10^{24}$\,Mx or even more.

Our method offers a quantitative investigation of highly resolved 
(reconstructed) intensity images, avoiding the ambiguities of the Stokes 
calibration procedure and allowing closely neighboring opposite polarities 
to be disentangled up to the spatial resolution achieved in two-dimensional 
brightness images.

\begin{acknowledgements}
Drs. J. S\'anchez Almeida, M. Sch\"ussler, and O. Steiner contributed 
helpful suggestions. Dr. P. S\"utterlin performed the observations
and image reconstruction, Dr. J. Hirzberger the power spectra.
An anonymous referee pointed out some critical problems. The 
computational work was performed under PV-WAVE (by Visual Numerics, 
Inc.). 
\end{acknowledgements}

\end{document}